\documentclass[proof]{pasj00}
\Received{$\langle$reception date$\rangle$}
\Accepted{$\langle$acception date$\rangle$}
\Published{$\langle$publication date$\rangle$}
\SetRunningHead{The SNR CTB37B with Suzaku}{R. Nakamura et al.}

\usepackage{times}

\begin{document}

\title{The Nature of a Cosmic-ray Accelerator, CTB37B, Observed \\
with Suzaku and Chandra}
\author{Ryoko \textsc{Nakamura}, Aya \textsc{Bamba}, Manabu \textsc{Ishida}}
\affil{
   Institute of Space and Astronautical Science/JAXA, \\
   3--1--1 Yoshinodai,Sagamihara, Kanagawa 229--8510}
   \email{nakamura@astro.isas.jaxa.jp, bamba@astro.isas.jaxa.jp, 
   ishida@astro.isas.jaxa.jp}
\author{Hiroshi \textsc{Nakajima}}
\affil{
   Department of Physics, Graduate School of Science, Kyoto University\\
   Kita-Shirakawa, Sakyo-ku, Kyoto 606--8502}
\author{Ryo \textsc{Yamazaki}}
\affil{
   Department of Physical Science, Hiroshima University, \\
   Higashi--Hiroshima, Hiroshima 739--8526}
\author{Yukikatsu \textsc{Terada}}
\affil{
   Department of Physics, Saitama University\\
   255 Shimo-Okubo, Sakura, Saitama 338--8570}
\and
\author{Gerd \textsc{P\"{u}hlhofer}, Stefan J. \textsc{Wagner}}
\affil{Landessternwarte, Universit\"{a}t Heidelberg, \\
K\"{o}nigstuhl, 69117 Heidelberg, Germany}

\KeyWords{acceleration of particles ---   ISM: individual~(CTB37B) ---
ISM: supernova remnants --- X-rays: ISM}

\maketitle

\begin{abstract}
 We report on Suzaku and Chandra observations of the young supernova
 remnant CTB37B, from which TeV $\gamma$-rays were detected by the
 H.E.S.S. Cherenkov telescope. The 80~ks Suzaku observation provided us
 with a clear image of diffuse emission and high-quality spectra. The
 spectra revealed that the diffuse emission is comprised of thermal and
 non-thermal components. The thermal component can be represented by an
 NEI model with a temperature, a pre-shock electron density and an age
 of 0.9$\pm0.2$ keV, 0.4$\pm0.1$~cm$^{-3}$ and 650$^{+2500}_{-300}$ yr,
 respectively. This suggests that the explosion of CTB37B occurred in a
 low-density space.
 A non-thermal power-law component was found from the southern region
 of CTB37B. Its photon index of $\sim$1.5 and a high roll-off energy
 ($\gtsim$15~keV) indicate efficient cosmic-ray acceleration. A comparison of this X-ray spectrum with the TeV
 $\gamma$-ray spectrum leads us to conclude that the TeV $\gamma$-ray
 emission seems to be powered by either multi-zone Inverse Compton
 scattering or the decay of neutral pions.
 The point source resolved by Chandra near the shell is probably
 associated with CTB37B, because of the common hydrogen column density
 with the diffuse thermal emission. Spectral and temporal
 characteristics suggest that this source is a new anomalous X-ray
 pulsar.
\end{abstract}

\section{Introduction}

Supernova Remnants (SNRs) are one of the most promising acceleration sites
of cosmic rays. In fact, ASCA detected synchrotron X-ray emission from
the shell of SN~1006, which unambiguously indicates the acceleration of
electrons up to $\sim$100~TeV \citep{1995Natur.378..255K}. Following
this discovery, the synchrotron X-ray emission has been discovered from
a shell of a few more SNRs, such as RX~J1713.7--3946
\citep{1997PASJ...49L...7K} and RCW~86 \citep{2000PASJ...52.1157B}.  On
the other hand, TeV $\gamma$-rays have also been detected from some
non-thermal shell-type SNRs. The radiation of TeV $\gamma$-ray is
explained by either (1)~Inverse-Compton scattering~(IC) of cosmic
microwave background photons by the same high energy electron giving
rise to the X-ray synchrotron emission or (2)~the decay of neutral pions
that are generated by collisions between high energy protons and dense
interstellar matter.
The ratio of fluxes between the TeV
$\gamma$-ray and the X-ray provides the magnetic field intensity as long
as one assumes that the TeV $\gamma$-ray is produced through the IC mechanism. Utilizing this characteristic,
\citet{2007PASJ...59S.199M} found that the TeV $\gamma$-ray from HESS~J1616$-$508 is likely the result of the proton acceleration, because the
non-detection of X-ray using the Suzaku XIS provides much weaker
magnetic field than the interstellar average. 

Although the evidence of particle acceleration has accumulated rapidly,
our knowledge is still limited on what sort of conditions are necessary
for SNRs to accelerate particles. A breakthrough may be brought about by
searching SNRs from which the TeV $\gamma$-ray emission is already
detected for thermal emission systematically, since the thermal emission
provides us with a lot of information on the environment such as
temperature, density, and age of the plasma.

CTB37B locates at ($l$, $b$) = ($\timeform{348D.7}$, $+\timeform{0D.3}$)
with a distance of $10.2\pm 3.5$~kpc \citep{1975A&A....45..239C}. This
region is one of the most active regions in our galaxy where star burst
activities, a number of shell structures probably associated with recent
SNRs \citep{1991ApJ...374..212K}, and OH maser sources
\citep{1996AJ....111.1651F} are detected in radio band.
TeV~$\gamma$-ray emission is also detected by the H.E.S.S. observation
\citep{2007A&A...472..489A}. In spite of the evidence of the high
activities in other wave bands, X-ray observations have been relatively
poor. Only ASCA \citep{1994PASJ...46L..37T} has detected a part of
CTB37B at the edge of the field of view of the Gas Imaging Spectrometer
(GIS) \citep{1996PASJ...48..157O,1996PASJ...48..171M} in the course of
the galactic plane survey (\cite{2001ApJS..134...77S}; Yamauchi et
al. 2008). Although the statistics are limited and the response of the
GIS is not qualified at the pointing position of CTB37B, Yamauchi et
al.~(2008) represents that the fit of a
power law to the GIS spectrum results in a steep photon index of
$\sim$4.1, whereas fit of an optically thin thermal plasma model
requires a high temperature of $\sim$1.6~keV. These results strongly
suggest that the X-ray spectrum is a mixture of a non-thermal power law
and an optically thin thermal plasma emission. In addition,
\citet{2008arXiv0803.0682H} resolved a bright point source located
near the shell of CTB37B from the diffuse emission by Chandra,
although its spectral parameters are not constrained very well
because of short exposure time.

In order to take an image and high quality spectra of CTB37B, we have
carried out an observation of CTB37B with Suzaku. We also refer to the
Chandra archival data to include the spatial structure and to compare
them to our Suzaku data. In \S~2, we present the observation log and
data reduction method. Image analysis is presented in \S~3. Spectrum
analysis and timing analysis are shown in \S~4 and \S~5,
respectively. We have really detected both the thermal and non-thermal
power-law components from CTB37B. Discussions are made in \S~6 on the
nature of the thermal and non-thermal component as well as the point
source. Finally we summarize our results in \S~7.

\section{Observation and Data Reduction}
\subsection{Suzaku Observation}
CTB37B was observed with Suzaku \citep{2007PASJ...59S...1M} during 2006
August 27--29. The nominal pointing position was (RA, Dec) =
($\timeform{17h13m57s}$, $-\timeform{38D12'15''}$, J2000). Suzaku is
equipped with two kinds of X-ray detectors; one is the Hard X-ray
Detector (HXD; \cite{2007PASJ...59S..35T}; \cite{2007PASJ...59S..53K}),
which is a non-imaging type detector and is sensitive in the 10-600~keV
band. The other is the X-ray Imaging Spectrometer (XIS;
\cite{2007PASJ...59S..23K}), which is an X-ray CCD camera mounted on the
focal plane of the X-Ray Telescope (XRT; \cite{2007PASJ...59S...9S}).
In total, there are four modules of the XIS, three of which are
Front-Illuminated~(FI) CCDs, which are hereafter referred to as XIS-0,
2, and 3, and the other one is a Back-Illuminated~(BI) CCD, which is
referred to as XIS-1. The XRT has a point-spread function (PSF) of a
Maltese-cross shape with a core radius of $\sim$15$''$ accompanied by an
outskirt extending a few arcmin. The half-power diameter (HPD) of each
telescope is $\sim$2~arcmin.
We concentrate on the XIS data in this paper because the HXD has no
imaging capability and hence there remains a large systematic error in
estimating the flux from CTB37B.

The XIS was operated in the normal full-frame clocking mode with neither
burst nor window options and SCI-off. The editing mode was 3~$\times$~3
in low and medium data rates and 5~$\times$~5 in high and super-high
data rates.  In analysis, we employed the data processed with the
revision 1.2 pipeline software, and used the HEADAS software (version
6.2) and XSPEC (version 11.3.2) for the data reduction and spectral
analysis, respectively. We applied the charge-transfer inefficiency
(CTI) correction by ourselves with the xispi software and CTI parameters
of 2006--08--23. After the screening of the data, the effective exposure
time of 80~ksec in total. The response matrix files (RMF) and ancillary
response files (ARF) were made using xisrmfgen and xissimarfgen
\citep{2007PASJ...59S.113I} version 2007--09--22 under the assumption
that the emissions are from point source.

\subsection{Chandra Observation}
Chandra observation was performed on the 2$^{th}$ February 2007 with the
Advanced CCD Imaging Spectrometer (ACIS). Chips I0, I1, I2, I3 S2 and S3
were used. The angular resolution is $\sim$0.5 arcsec
which correspond to the CCD pixel size. The data
reduction and analysis were made using the Chandra Interactive Analysis
of Observations (CIAO version 3.4, CALDB version 3.3.0). The total
exposure time is 26~ksec after screening the data.

\section{Image Analysis}

\begin{figure}[htpb]
 \begin{center}
\FigureFile(60mm,35mm){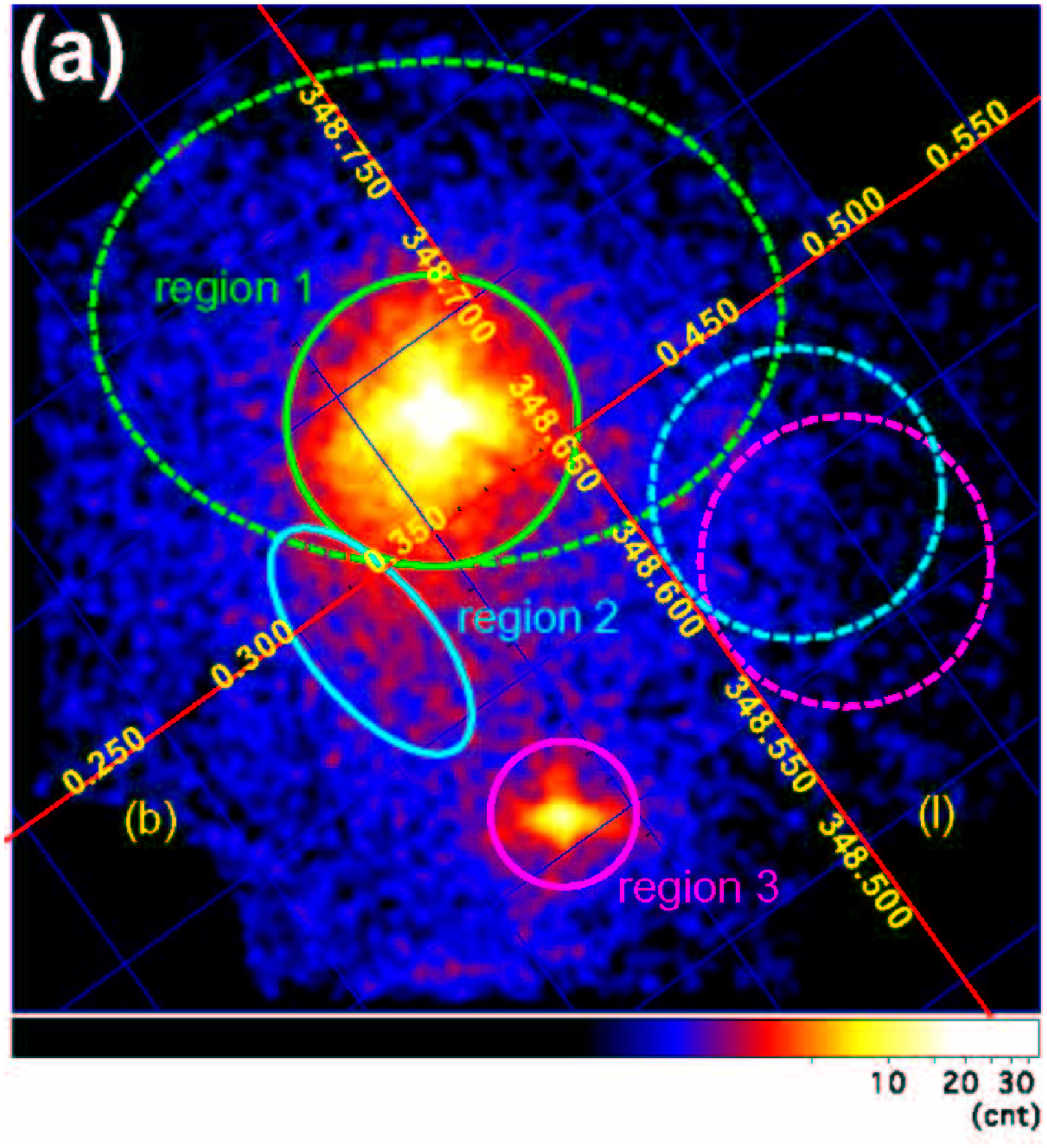} \FigureFile(60mm,35mm){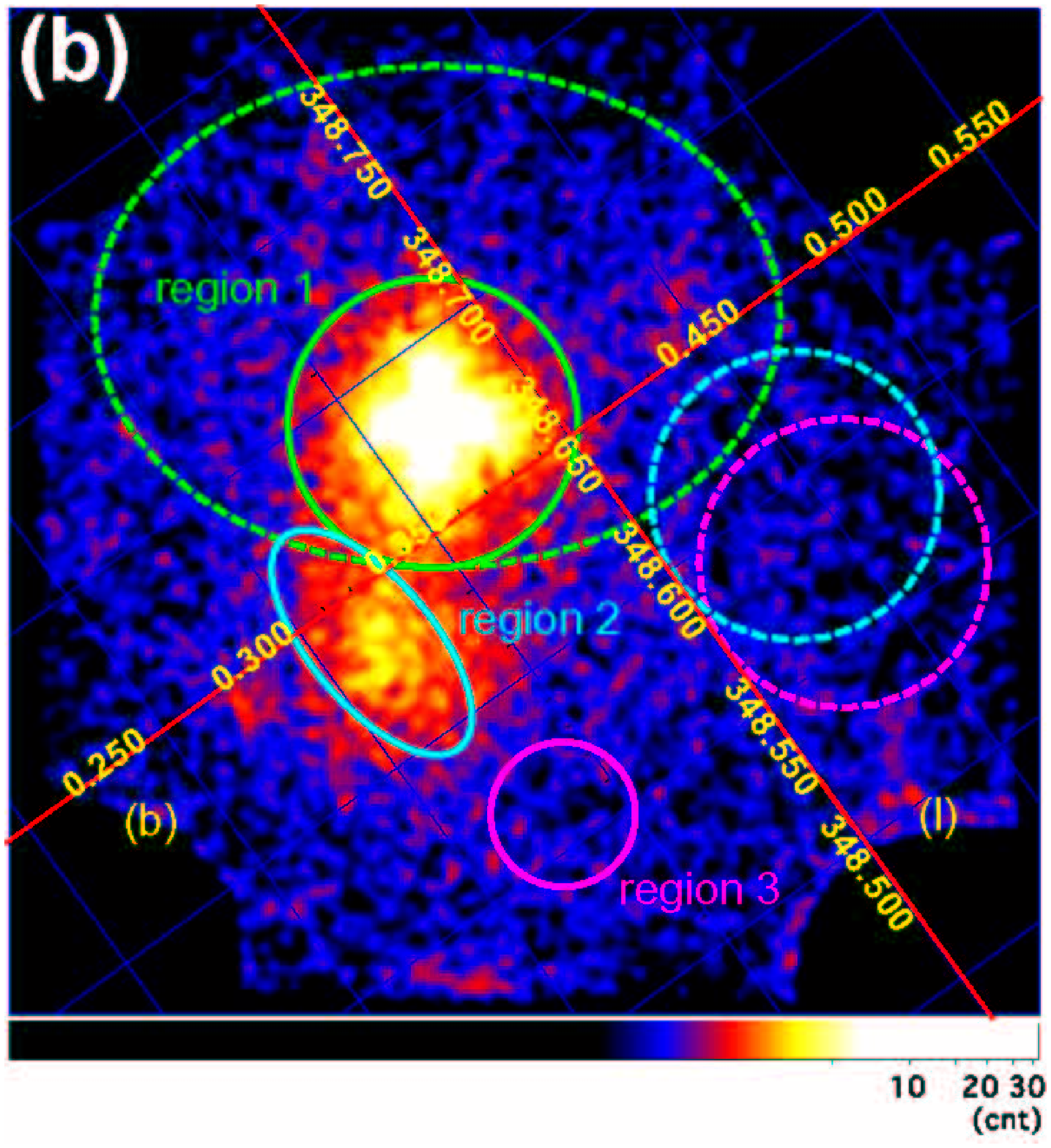}
\FigureFile(60mm,35mm){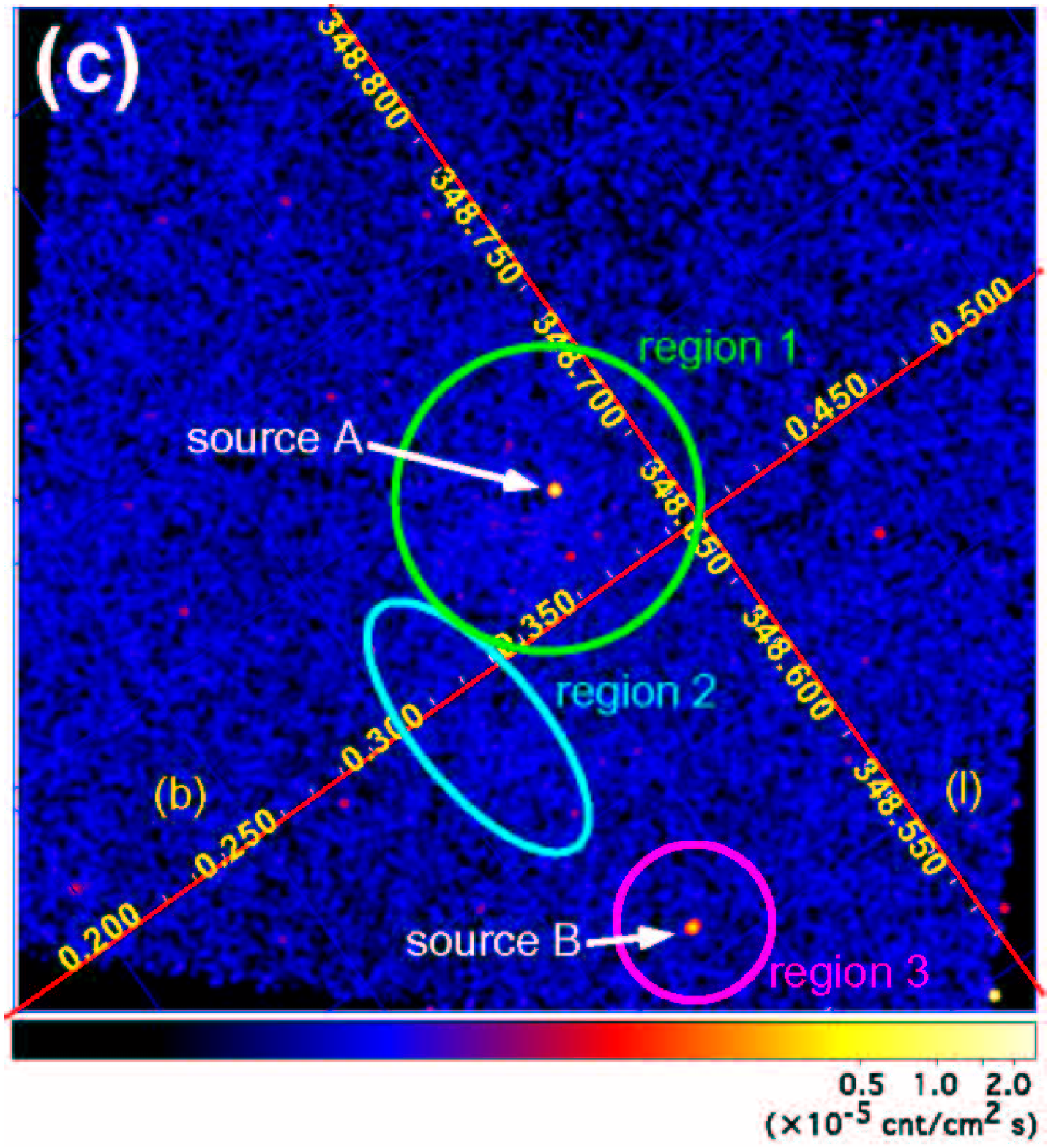}\\
\FigureFile(60mm,35mm){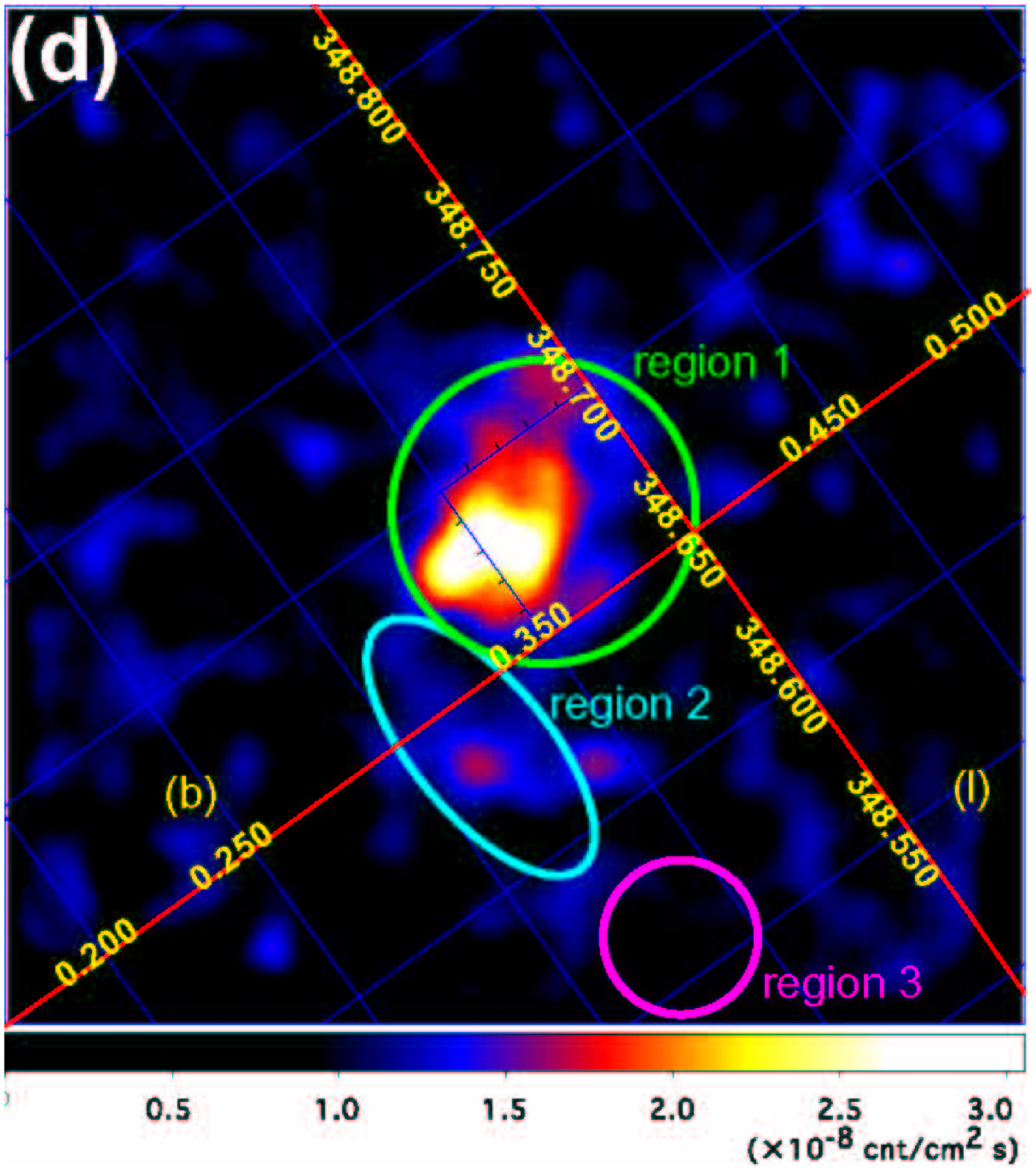} \FigureFile(60mm,35mm){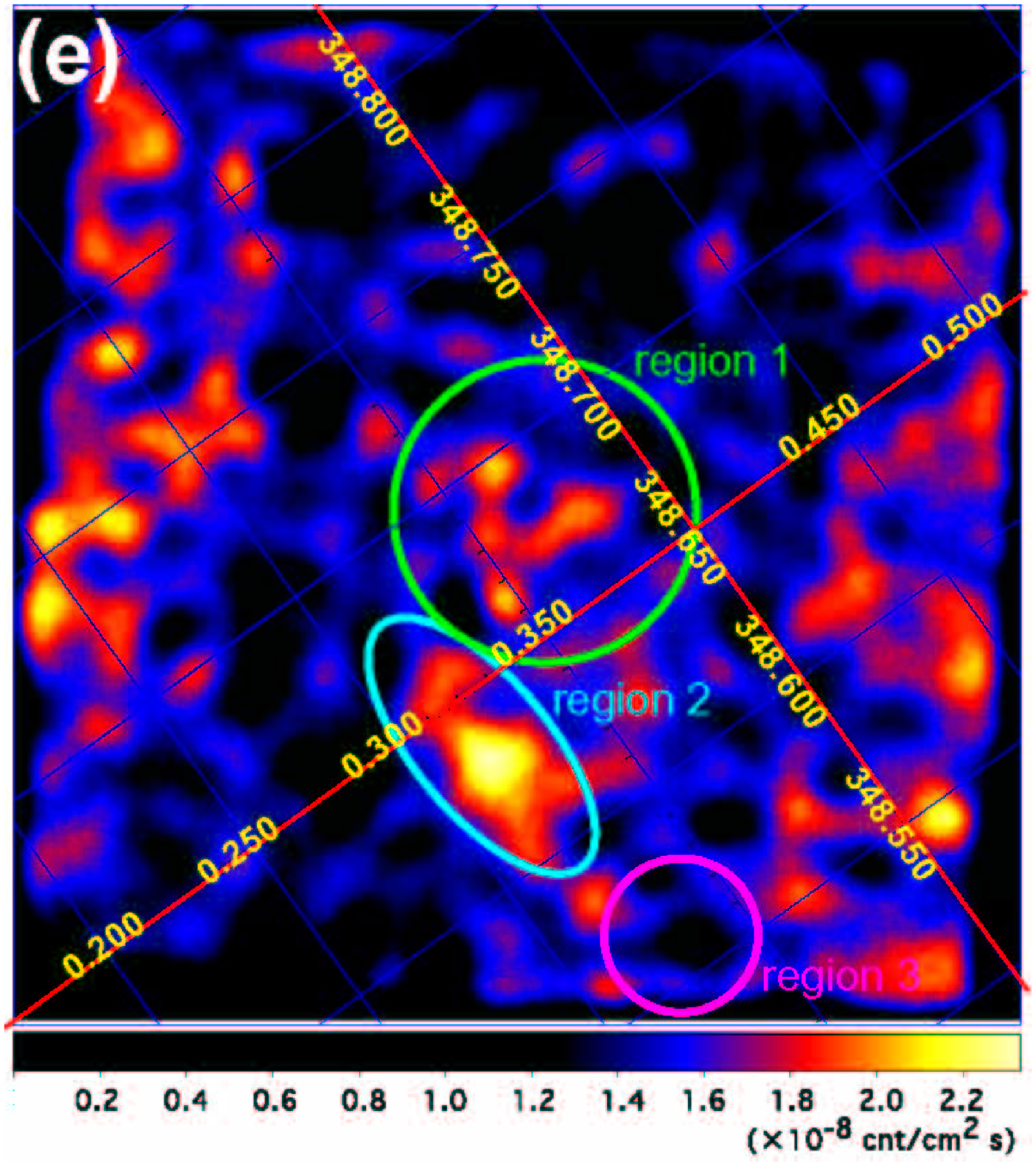}
 \end{center}
\caption{Images of CTB37B in the galactic coordinates. Panels (a) and
 (b) are Suzaku images in 0.3--3.0~keV and 3.0--10.0~keV, respectively,
 which are smoothed with a Gaussian with $\sigma = 12$~arcsec. Panel (c)
 is the Chandra image in the 0.3--10.0~keV band being smoothed with a
 Gaussian with $\sigma = 6$~arcsec. Panels (d) and (e) are the Chandra
 images in 0.3--3.0~keV and 3.0--10.0~keV, respectively. After removing
 point sources, we smoothed them with a Gaussian with a $\sigma$ of
 40~arcsec. Solid circles in green, blue, and magenta are the
 integration region of source photons which are named as region 1
 through 3 in this order. Region 1 is a circle with a radius of
 $\timeform{2.'6}$, region 2 is an ellipse with a size of
 $\timeform{1.'1}\times\timeform{2.'5}$, and region 3 is a circle with a
 radius of $\timeform{1.'3}$. The dashed regions are corresponding
 background-integration regions.}
\label{fig:image_xray}
\end{figure}
\subsection{Suzaku Images}
Fig.~\ref{fig:image_xray} shows Suzaku XIS images in
0.3--3.0~keV and 3.0--10.0~keV. They are created by combining those from
all the four XIS modules and  
smoothed with a Gaussian with $\sigma =
12$~arcsec, which is close to the XRT core size and
effective in highlighting the diffuse emission. The
source that locates at ($l$, $b$) $\simeq$ ($\timeform{348D.68}$,
$\timeform{0D.37}$) appears as the brightest source both in the soft and
hard bands.
Another source extending to the south of the brightest source, at ($l$,
$b$) $\simeq$ ($\timeform{348D.63}$, $\timeform{0D.32}$) seems to be a
diffuse source and manifests itself only in the band above 3~keV. In
addition to these sources, a point source is detected at ($l$, $b$)
$\simeq$ ($\timeform{348D.56}$, $\timeform{0D.33}$) in the band below
3~keV. The sky position is consistent with that of the point source 1RXS
J171354.4$-$381740 listed in the ROSAT Bright Star Catalogue
\citep{1999A&A...349..389V}. In order to investigate these sources
separately, we defined the following photon-integration regions (see
Fig.~\ref{fig:image_xray}) for the spectral analysis. Region 1 is the
green circle with a radius of $\timeform{2'.6}$ centered at the
intensity peak of the brightest source. Region 2 is the blue ellipse
with a major and minor axis of $\timeform{2.'5}$ and
$\timeform{1.'1}$, respectively, which is centered at the second
diffuse source. Region 3 is the circle colored in magenta with a radius
of $\timeform{1'.3}$. The other three regions with the same colors but
with dashed lines define those collecting the background events.  We set
these background regions by taking into account the telescope
vignetting.

\subsection{Chandra Images}
Fig.~\ref{fig:image_xray}(c) shows the Chandra image in the
0.3--10.0~keV band corrected for the telescope vignetting and smoothed
with a Gaussian with $\sigma = 6$~arcsec to see the
point sources clearly. In total, 18 sources are detected above
5$\sigma$ confidence level \citep{2008arXiv0803.0682H}.  The panels (d)
and (e) are images in 0.3--3.0~keV and 3.0--10.0~keV, respectively,
after removing the point sources. They are smoothed with a Gaussian with
a $\sigma =$ 40~arcsec to highlight
the diffuse emission. The definition of the colored regions is same as
those drawn in the Suzaku images. The brightest source in region 1 is a
point source, which is defined as source~A, locating at ($l$, $b$) =
($\timeform{348D.681}$, $\timeform{0D.371}$).  Although most of the flux
from region~1 originates from the point source, it is revealed from
Fig.~\ref{fig:image_xray}(d) that region~1 is accompanied by a diffuse
emission.
These two components are mixed in the Suzaku images. As expected from
the Suzaku images in Fig.~\ref{fig:image_xray}(a) and (b), region~2 is
brighter than region 1 in the band above 3~keV. In region 3, there is
the second bright source~(defined as source~B, which is
1RXS~J171354.4$-$381740) locating at ($l$, $b$) =
($\timeform{348D.561}$, $\timeform{0D.332}$), and no diffuse emission is
associated in 0.3--3.0~keV and 3.0--10.0~keV
band. Table~\ref{tab:source_count} summarizes the
background subtracted source count rates separately for point and
diffuse sources in region 1, region 2 and region 3.

 \begin{table}
 \begin{center}
 \caption{Count rates of sources and diffuse emission of Chandra data.}
\label{tab:source_count}
  \begin{tabular}{lccc}
   \hline\hline
   \multicolumn{1}{c}{Region} & Energy band  & Point source count rate \footnotemark[*]&
   Diffuse emission count rate \footnotemark[*]\\
 & & [10$^{-2}$ counts s$^{-1}$]
 & [10$^{-2}$ counts s$^{-1}$] \\ \hline
   region 1 & 0.3--3.0~keV  & 2.7~$\pm$0.1 & 3.3~$\pm$0.2 \\
                             & 3.0--10.0~keV & 2.4~$\pm$0.1 & 0.64~$\pm$0.26 \\
                             & total & 5.0~$\pm$0.1 & 3.9~$\pm$0.4 \\
   region 2 & 0.3--3.0~keV  & --- & 0.43~$\pm$0.15 \\
                             & 3.0--10.0~keV & --- & 0.93~$\pm$0.19\\
                             & total & --- & 1.4~$\pm$0.2\\
   region 3 & 0.3--3.0~keV  & 2.3~$\pm$0.1 & --- \\
                             & 3.0--10.0~keV & 0.024~$\pm$0.021 & --- \\
                             & total & 2.3~$\pm$0.1 & ---\\
   \hline\hline
\multicolumn{4}{@{}l@{}}{\hbox to 0pt{\parbox{80mm}{\footnotesize
 Note --- All errors are at 1$\sigma$ confidence level.
}\hss}}
 \end{tabular}
\end{center}
\end{table}

\subsection{Correlation with Other Energy Band}
\begin{figure}[htpb]
 \begin{center}
\FigureFile(80mm,50mm){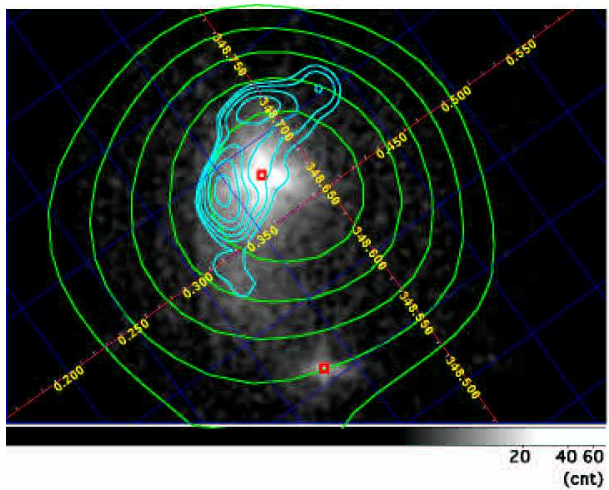}
 \end{center}
\caption{Gray scale Suzaku X-ray image with radio and TeV~$\gamma$-ray
 contours (blue and green, respectively) overlaid. The
 blue radio contours are in logarithmic scale from 0.016 to 1.3 Jy/beam.
 The green TeV contours are in linear scale from 30 to 50
 count/0.9~arcmin$^{2}$ stepped by 4. Red boxes are source A
 and B, and green cross is the center of TeV $\gamma$-ray emission.}
\label{fig:image_other}
\end{figure}

Fig.~\ref{fig:image_other} shows brightness contours of radio at 1.4~GHz
in blue and of TeV $\gamma$-ray with H.E.S.S. in green, overlaid on the
gray scale image of the Suzaku in the 0.3--10.0~keV band. 
Source~A and B resolved by Chandra are
represented by the filled red boxes.
The radio image is taken from the NRAO VLA Sky Survey (NVSS)
database\footnote{http://www.cv.nrao.edu/nvss/}
\citep{1998AJ....115.1693C}. The X-ray emission well conforms with the
shell in radio. Particularly, the diffuse X-ray source detected by
Suzaku in region 2 is associated with the southern radio sub-peak. On
the other hand, the peak of TeV $\gamma$-ray emission and source~A
seems to be separated from the radio shell.  Note, however, that, due to
limited spatial resolution of H.E.S.S, the apparent TeV source
morphology is consistent with that of the radio shell whose radius is
$\timeform{4.'5}$ \citep{2008arXiv0803.0682H}.

\section{Spectral Analysis}
In this section, we present results of spectral analysis of the three
regions described above. We adopt the metal composition of
\citet{1989GeCoA..53..197A} as the solar abundance. Spectral fits are
carried out with XSPEC. We always adopt an ancillary response file (ARF)
for a point source, since the sizes of region 1 and 2 are so small that
the resultant spectral parameters including the flux will differ only by
$\sim$1\% from the case if we take into account the spatial extent. The
errors quoted are always at the 90\% confidence level.

\subsection{Region 1}
\begin{figure}[htpb]
 \begin{center}
\FigureFile(80mm,50mm){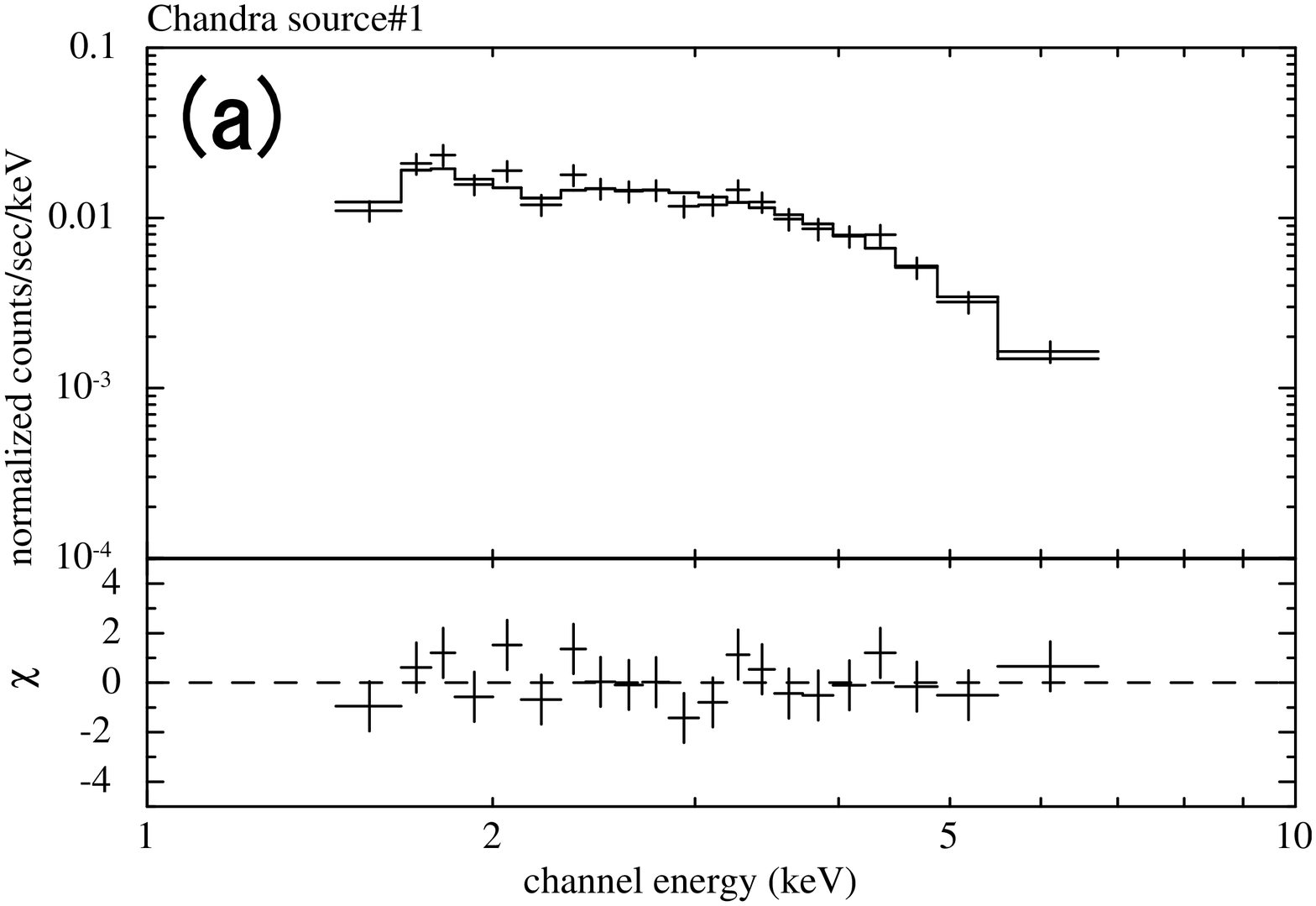}
\FigureFile(80mm,50mm){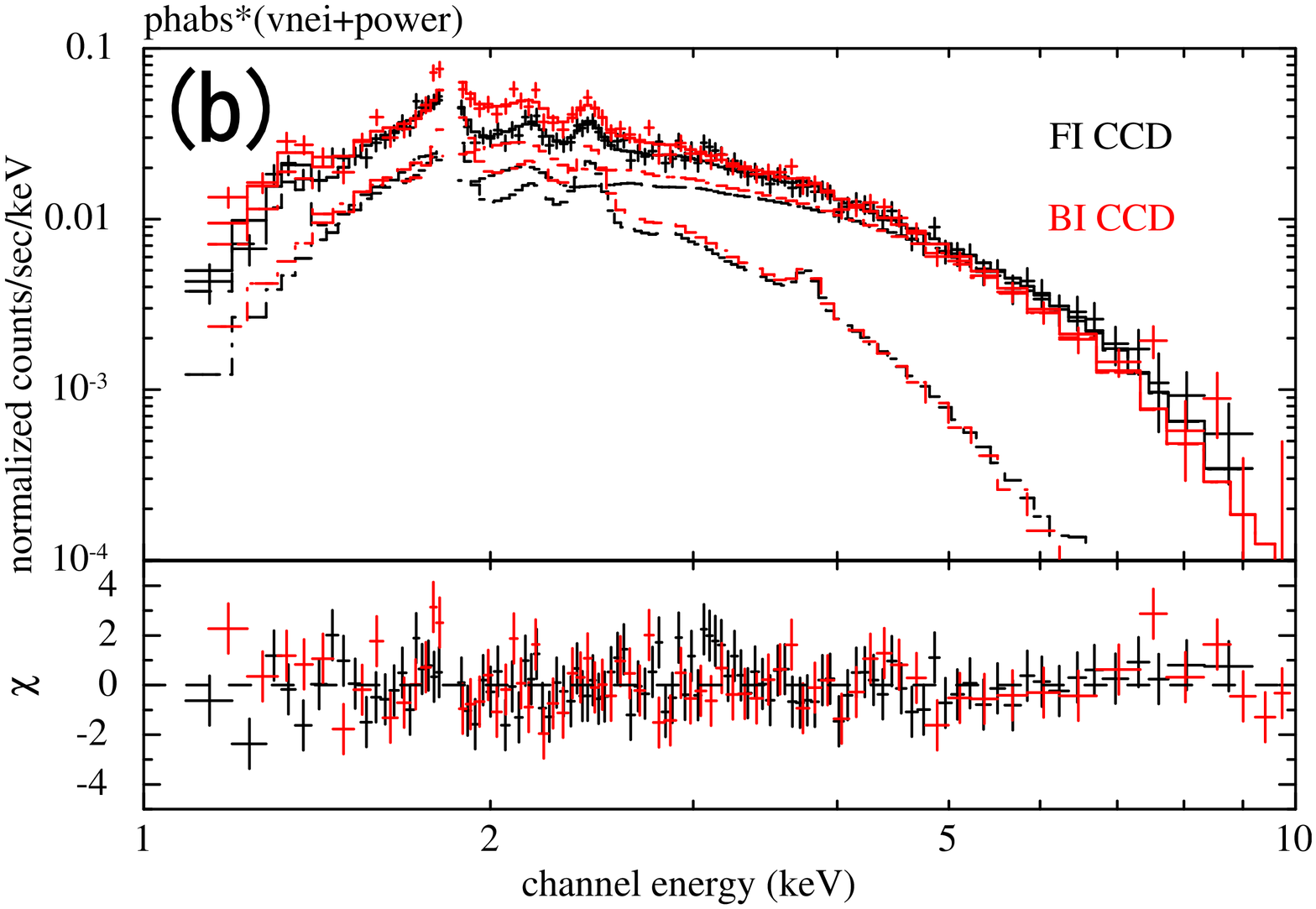}
 \end{center}
\caption{(a) Chandra spectrum integrated from a circular region with a
 radius of $3''$ centered on source~A. The best-fit model represented
 by a power law (histogram) is overlaid. (b) Suzaku spectra of region~1
 from the FI-CCDs (black) and the BI-CCD (red) with a model composed of
 a ``vnei'' and a power law model. The best-fit parameters are
 summarized in table~\ref{tab:param_1}. \label{fig:region1_spec_fit}}
\end{figure}

The Chandra spectrum of source~A is shown in
Fig.~\ref{fig:region1_spec_fit}(a). In extracting this spectrum, we took
a circular integration region with a radius of $3''$ centered at the
source. We made no background subtraction. Since there is no apparent
emission lines, we attempted to fit a power-law model undergoing
photoelectric absorption (``phabs'' model in XSPEC) to the spectrum.
The best-fit model is overlaid in the upper panel of
Fig.~\ref{fig:region1_spec_fit}(a) as the histogram. The best-fit
parameters are summarized in table~\ref{tab:param_1}. The photon index,
the hydrogen column density, and the intrinsic flux in the 2.0--10.0~keV
band are $3.2^{+0.4}_{-0.3}$, 4.0~($\pm0.6)\times10^{22}$~cm$^{2}$, and
$1.8~(\pm0.2)\times10^{-12}$~ergs~cm$^{-2}$s$^{-1}$, respectively, which
are consistent with those of \citet{2008arXiv0803.0682H}.

Fig.~\ref{fig:region1_spec_fit}(b) is the background-subtracted spectrum
of Suzaku region 1. The black and red crosses represent the data points
from the sum of the FI CCDs and those of the BI CCD, respectively.
Although there is no sign of Fe K$\alpha$ line in the 6--7~keV band, we
have obviously detected K$\alpha$ emission lines from He-like Mg
(1.34~keV), Si (1.86~keV), and S (2.46~keV) as well as a K$\beta$
emission line from He-like Si (2.18~keV). This means that the spectrum
includes a thermal emission component. We thus tried to fit the Suzaku
spectrum with a model composed of a power law representing source~A and
a non-equilibrium collisional ionization plasma emission model (``vnei''
model in XSPEC;
\cite{2001ApJ...548..820B,1983ApJS...51..115H,1994ApJ...429..710B,1995ApJ...438L.115L})
undergoing photoelectric absorption with a common $N_{\rm H}$. In the
fitting, we set abundances of Mg, Si, and S free to
vary. The other abundances are fixed as solar
abundance. The best-fit parameters are summarized in
Table~\ref{tab:param_1}, and the best-fit models are displayed in
Fig.~\ref{fig:region1_spec_fit}(b). The fact that no iron K$\alpha$
emission line was detected can be
attributed  that the
non-thermal component dominates the spectrum in the energy band above
3~keV. The reduced $\chi^2$ of 1.06 implies that the fit is acceptable
at the 90\% confidence level. The temperature and the ionization
parameter of the ``vnei'' component are obtained to be
$kT=0.89^{+0.22}_{-0.17}$~keV and $n_{\rm e}t\,{\rm [cm^{-3}s]} =
3.5^{+13}_{-1.1}\times10^{10}$, respectively. On the other hand, the
photon index of the power-law model is $\Gamma =3.0\pm 0.2$ and the
intrinsic flux is $3.3^{+0.3}_{-0.4}\times10^{-12}$
ergs~cm$^{-2}$s$^{-1}$. The photon index is consistent between Suzaku
and Chandra, whereas the flux with Suzaku seems to be greater than that
with Chandra by a factor of $\sim$1.8. We remark that, since there is no
sign of emission lines in the Chandra spectrum of source~A, the
thermal component should be extended.

\begin{table}
\begin{center}
\caption{Best-fit parameters of the region 1 spectra}
\label{tab:param_1}
 \begin{tabular}{lcc}
 \hline \hline
 \multicolumn{1}{c}{Parameters}       & Chandra source~A & Suzaku \\
 \hline
 Power Law & & \\
  \hspace{1pc}Photon Index & 3.2$^{+0.4}_{-0.3}$ & 3.0~$\pm$0.2 \\
  \hspace{1pc}Intrinsic Flux \footnotemark[a] & 1.8~$\pm$0.2 & 3.3$^{+0.3}_{-0.4}$ \\
 VNEI & & \\
  \hspace{1pc}Temperature \ [keV] & $\cdots$ & 0.89$^{+0.21}_{-0.17}$\\
  \hspace{1pc}abundance \footnotemark[b] \ Mg & $\cdots$ & 0.61$^{+0.31}_{-0.19}$\\ 
   \hspace{5.7pc}Si & $\cdots$ & 0.40$^{+0.21}_{-0.14}$ \\
   \hspace{5.8pc}S & $\cdots$ & 1.0$\pm$0.6 \\
 {\hspace{1pc}\it n$_e$t} \footnotemark[c] & $\cdots$ & 3.5$^{+13}_{-1.1}$\\
 \hspace{1pc}{\it E.M.} \footnotemark[d] & $\cdots$ & 2.1$^{+1.6}_{-1.0}$\\
 $N_{\rm H}$ \footnotemark[e] & 4.0~$\pm0.6$ & 3.6$^{+0.4}_{-0.2}$\\ 
$\chi$ $^2$/d.o.f & 14.6/18 & 176.1/166\\
\hline \hline
 \multicolumn{3}{@{}l@{}}{\hbox to 0pt{\parbox{85mm}{\footnotesize
     \footnotemark[a] Flux in the 2.0--10.0~keV band in
 the unit of 10$^{-12}$ \ ergs \ cm$^{-2}$ \ s$^{-1}$.\\
     \footnotemark[b] Abundance ratio relative to the
 solar value (Anders \& Grevesse, 1989).\\
     \footnotemark[c] Ionization time--scale in the
 unit of 10$^{10}$ \ s \ cm$^{-3}$ , where {\it n$_e$} and {\it t} are
 the electron density and age of the plasma.\\
    \footnotemark[d] Emission measure {\it E.M.} =
 $\int n_{e}n_{H}dV \simeq n_{e}^{2}V$ in the unit of 10$^{58}$ \
 cm$^{-3}$, where $ n_{e}$ and $V$ are the electron density and the
  plasma volume. The distance to CTB37B is assumed to be 10.2~kpc 
  \citep{1975A&A....45..239C}\\
    \footnotemark[e] Absorption column density in the unit of
 10$^{22}$ \ cm$^{-2}$.\\
     }\hss}}
 \end{tabular}
\end{center}
\end{table}

\subsection{Region 2}
\begin{figure}[htpb]
 \begin{center}
\FigureFile(80mm,50mm){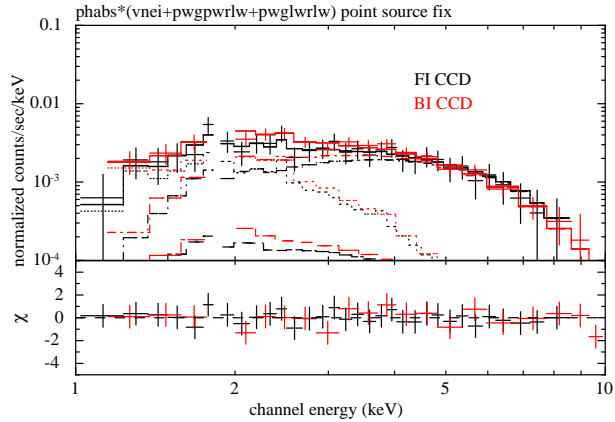}
 \end{center}
\caption{The XIS spectra of region~2 with a model composed of a ``vnei''
 and two power-law components, of which one represents contamination
 from source~A. The normalization of ``vnei'' model and all parameters
 of the other power law are set free to vary.
 \label{fig:region2_spec_fit}}
\end{figure}

The Suzaku spectra of region 2 together with best-fit model and residual
are shown in Fig.~\ref{fig:region2_spec_fit}.
The basic features such as the He-like Si K$\alpha$ emission line and no
apparent sign of Fe emission line are similar to those in region~1. In
addition to the thermal and non-thermal components, we need to take into
account possible contamination from source~A which is brightest in
region~1. In fitting the region 2 spectra, we thus first tried a ``vnei
+ power~law~(1) + power~law~(2)'' model, where the power~law~(1)
accounts for the source~A contamination, and the power~law~(2)
represents a non-thermal component dominating the high energy band image
in region~2 (Fig.~\ref{fig:image_xray}).
The vnei component, on the other hand, represents the
contamination from region 1, and intrinsic thermal emission from region
2 if any. Note that the region~2 spectra are statistically poorer than
those of region~1. Hence we have fixed the temperature, the abundances
of Mg, Si, and S, the ionization parameter of the vnei
component and the photon index of power~law~(1) at the best-fit values
obtained in the region~1 fit, which are summarized in
table~\ref{tab:param_1}. The flux between 2.0--10.0~keV of power~law~(1)
is fixed at 3.3$\times10^{-14}$~ergs~cm$^{-2}$~s$^{-1}$, which is
$\sim$1\% of the source~A flux, on the basis of the vignetting function
\citep{2007PASJ...59S...9S}. As a result, the free parameters are the
hydrogen column density, the normalization of the vnei
component, and all parameters of power~law~(2).
The result of the fit is shown in Fig.~\ref{fig:region2_spec_fit}, and
the best-fit parameters are summarized in
table~\ref{tab:param_2}. The $E.M.$ of the vnei component is
$\sim$~11~(4-20)~\% of the region 1 best-fit
value. On the other hand, we simulated the
contamination of the vnei component from region 1 using smoothed Chandra
image, and found the contamination is
$\sim$~7\%. The thermal component apparent in region 2
spectra can therefore be entirely regarded as the contamination from
region 1, and the upper limit of the $E.M.$ intrinsic to region 2 is 13\%
of that of region 1, or 2.7$\times10^{57}$~cm$^{-3}$. The photon index
of power~law~(2) results in $\Gamma = 1.5~(\pm0.4)$ with the reduced
$\chi^2$ of 0.36. It is remarkable that the X-ray photon index $\Gamma =
1.5$ is consistent with the standard radio energy index of non-thermal
SNRs $\alpha = 0.5$.

We next replaced the power-law~(2)
component by an ``srcut'' model, which simulates a synchrotron spectrum
from an exponentially cut off power-law distribution of electrons in a
homogeneous magnetic field
\citep{1998ApJ...493..375R,1999ApJ...525..368R}.
According to the Green's
catalogue\footnote{http://www.mrao.cam.ac.uk/surveys/snrs/}, the radio
spectral index~($\alpha$) is 0.3 with a flux at 1~GHz of 26~Jy. This
small index, however, is probably due to contamination of thermal
emission. A similar situation has been reported for 30 Dor C
\citep{2004ApJ...611..881S}. We thus fixed $\alpha$ at 0.5, which is the
typical value of the SNRs in the radio band, and set the flux at 1~GHz
free to vary. As a result, the normalization of the srcut model is
obtained to be 1.43~mJy at 1~GHz. This is much smaller than the radio
flux 26~Jy at 1~GHz. Note, however, that this radio
normalization is an integration of the entire radio image of CTB37B,
part of which is, however, spilt out of regions 1 and 2. Moreover, the
radio flux encompassed in region 2 is only $\approx$2\% of that in
region 1. We therefore estimate the radio flux within region 2 to be
$\approx$1\% of the total, or $\approx 300$~mJy at 1~GHz. Even after
this correction, simple extrapolation of the srcut model well fit to the
X-ray spectra to the radio band is much smaller than the observed radio
flux. We guess that the flux in the radio band is dominated by thermal
emission.
The resultant normalization (emission measure) of the vnei component
does not change within the error. The reduced $\chi^2$ value is nearly
the same as that of the power-law fit. The lower limit of the roll-off
energy is obtained to be 14.8~keV. We confirmed that the roll-off energy
does not change drastically if we varied $\alpha$ in the range
0.3--0.7. Based on these results, it is possible to interpret that the
spectrum of the non-thermal component extends
from X-ray to radio with an energy
index of 0.5 in the radio band.

 \begin{table}
 \begin{center}
\caption{Best-fit parameters of the region 2 spectrum}
\label{tab:param_2}
  \begin{tabular}{lcc}
   \hline\hline
   \multicolumn{1}{c}{Parameters}       & VNEI + Powerlaw + Powerlaw &
   VNEI + Powerlaw + srcut \\
   \hline
 VNEI & & \\
  \hspace{1pc}Temperature \ [keV] & 0.89~(fix) & 0.89~(fix)\\
  \hspace{1pc}abundance \footnotemark[a] \ Mg & 0.61~(fix) & 0.61~(fix)\\
   \hspace{5.7pc}Si & 0.40~(fix) & 0.40~(fix) \\
   \hspace{5.8pc}S & 1.0~(fix) & 1.0~(fix) \\
 {\hspace{1pc}\it n$_e$t} \footnotemark[b] & 3.5~(fix) & 3.5~(fix) \\
 \hspace{1pc}{\it E.M.} \footnotemark[c] & 0.23$^{+0.18}_{-0.15}$ & 0.23$^{+0.14}_{-
0.15}$\\
 Power Law~(1) & & \\
  \hspace{1pc}Photon Index & 3.0~(fix) & 3.0~(fix) \\
  \hspace{1pc}Intrinsic Flux \footnotemark[d] & 0.033~(fix) &
   0.033~(fix) \\
 Power Law~(2) & & \\
  \hspace{1pc}Photon Index & 1.5~$\pm$0.4 & $\cdots$ \\
  \hspace{1pc}Intrinsic Flux \footnotemark[d] & 0.78$^{+0.07}_{-0.08}$ &
   $\cdots$ \\
srcut & & \\
\hspace{1pc}alpha & $\cdots$ & 0.5~(fix)  \\
\hspace{1pc}roll-off E [keV] & $\cdots$ & \textgreater 14.8 \\
\hspace{1pc}Normalization $^d$ & $\cdots$ & 1.4 \\
\hspace{1pc}Intrinsic Flux $^e$ & $\cdots$ & 0.78 \\
 phabs $N_{\rm H}$ \footnotemark[f] & 3.5$^{+0.5}_{-0.7}$ & 3.5$^{+0.5}_{-0.7}$\\ 
$\chi$ $^2$/d.o.f & 17.5/48 & 17.5/48\\
\hline \hline
 \multicolumn{3}{@{}l@{}}{\hbox to 0pt{\parbox{130mm}{\footnotesize
     \footnotemark[a] Abundance ratio relative to the
 solar value (Anders \& Grevesse, 1989).\\
     \footnotemark[b] Ionization time--scale in the
 unit of 10$^{10}$ \ s \ cm$^{-3}$ , where {\it n$_e$t} and {\it t} are
 the electron density and age of the plasma.\\
    \footnotemark[c] Emission measure {\it E.M.} =
 $\int n_{e}n_{H}dV \simeq n_{e}^{2}V$ in the unit of 10$^{58}$ \
 cm$^{-3}$, where $ n_{e}$ and $V$ are the electron density and the
   plasma volume. The distance to CTB37B is assumed to be 10.2~kpc 
  \citep{1975A&A....45..239C}\\
   \footnotemark[d] Radio flux at 1 GHz in the
   unit of 10$^{-3}$ Jy.\\
   \footnotemark[e] Flux in the 2.0--10.0~keV band in
 the unit of 10$^{-12}$ \ ergs \ cm$^{-2}$ \ s$^{-1}$.\\
    \footnotemark[f] Absorption column in the unit of
 10$^{22}$ \ cm$^{-2}$.\\
     }\hss}}
   \end{tabular}
 \end{center}
 \end{table}

\subsection{Region 3}
\begin{figure}[htpb]
 \begin{center}
\FigureFile(80mm,50mm){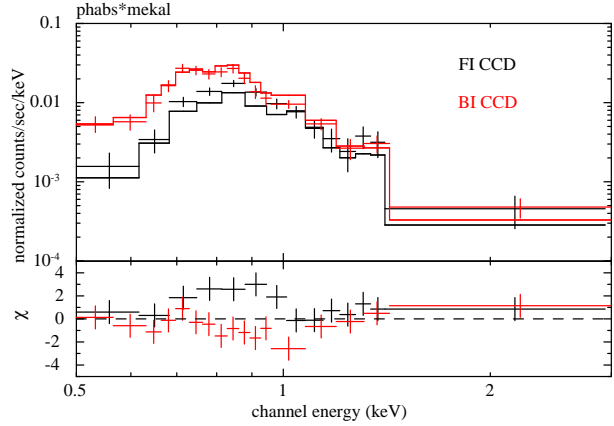}
 \end{center}
\caption{The XIS spectra of region 3 with the MEKAL model.}
\label{fig:region3_spec_fit}
\end{figure}

Fig.~\ref{fig:region3_spec_fit} shows the background-subtracted spectra
of region 3.
As indicated by the images in Fig.~\ref{fig:image_xray}(a), X-ray flux
is detected only below $\sim$3~keV.  Since the absorption is apparently
weak and there is Fe-L hump in the 0.7-0.9~keV band, this source seems
to be a foreground point source, probably an active star. We thus
adopted a model composed of a thin thermal collisional equilibrium
plasma emission model (``mekal'' model in XSPEC;
\cite{1985A&AS...62..197M,1986A&AS...65..511M,1995ApJ...438L.115L,1996uxsa.conf..411K})
multiplied by photoelectric absorption, and fitted this model to the
spectra in the 0.5--2.0~keV band. The result is shown in
Fig.~\ref{fig:region3_spec_fit}, and the best-fit parameters are listed
in Table~\ref{tab:param_3}. Note that the fit residuals exhibit
different behavior in the 0.7--1.0~keV band between the FI and BI
CCDs. This is probably attributed to calibration uncertainty.

\begin{table}[htpb]
\caption{Best-fit parameters of the region 3 spectrum}\label{tab:param_3}
\begin{center}
\begin{tabular}{ll}\hline\hline
MEKAL & \\
\hspace{2pc}Temperature [keV] & 0.46$^{+0.03}_{-0.05}$ \\
\hspace{2pc}Abundance \footnotemark[a] & 0.20$^{+0.12}_{-0.07}$ \\
\hspace{2pc}Normalization \footnotemark[b] [cm$^{-3}$]  & 2.5$^{+1.4}_{-0.6}$$\times10^{-4}$ \\
phabs & \\
\hspace{2pc}$N_{\rm H}$ \footnotemark[c] & \textless 0.046 \\ 
$\chi$ $^2$/d.o.f & 52.9/27 \\
\hline\hline
 \multicolumn{2}{@{}l@{}}{\hbox to 0pt{\parbox{70mm}{\footnotesize
     \footnotemark[a] Abundance ratio relative to the
 solar value (Anders \& Grevesse, 1989).\\
     \footnotemark[b] Normalization is
 $\frac{10^{-14}}{4\pi D_{A}^{2}} \int n_{e}n_{H}dV$ 
where $D_{A}$
 is the angular size distance to the source~(cm), $n_{e}$ and $V$ are the electron density and the plasma volume.\\
    \footnotemark[c] Absorption column in the unit of
 10$^{22}$ \ cm$^{-2}$.\\}}}
\end{tabular}
\end{center}
\end{table}

\section{Timing Analysis}

In order to understand the nature of the point sources, we carried out
timing analysis.

\subsection{Source A}
\begin{figure}[htpb]
 \begin{center}
\FigureFile(80mm,50mm){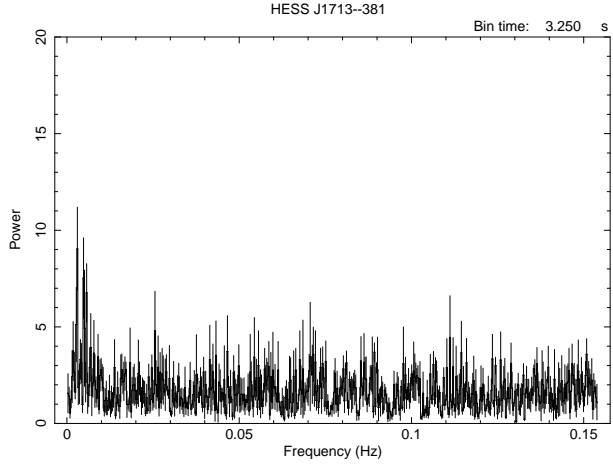}
 \end{center}
\caption{Power spectrum of source~A in the 0.3--10.0~keV band with
  the time bin size of 3.2~sec.}
\label{fig:source1_power}
\end{figure}

We searched the Chandra data for a pulsation from source~A.
Unfortunately, source~A locates close to the chip boundaries of the
ACIS-I, and is affected by the instrumental dithering effects. In fact,
the source drops into the boundary every 1000~s, during which the source
count diminishes significantly. Removing these time intervals with a
Good Time Interval filtering, we made a light curve and search for a
pulsation. The power spectrum on the basis of a 3.2~s binning light curve
in the 0.3--10~keV band is shown in Fig.~\ref{fig:source1_power}.
We do not detect any pulsation in the period range between 6.6--3000~s.

\subsection{Source B}
\begin{figure}[htpb]
 \begin{center}
\FigureFile(80mm,50mm){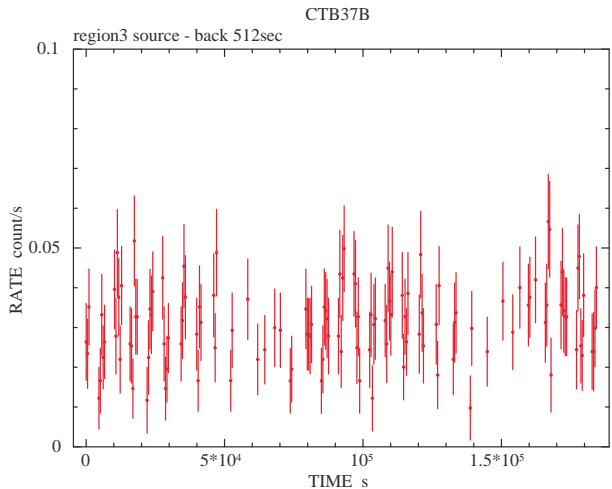}
 \end{center}
\caption{The light curve of region 3 in the 0.5--2.0~keV band. The time
 bin size is 512~s.}
\label{fig:region3_lc}
\end{figure}

We created a light curve of source~B in the 0.5--2.0~keV band with
Suzaku, which is shown in Fig.~\ref{fig:region3_lc}.
The time bin size is 512~s. Although we detected no drastic flare
event, the light curve seems to show flickering. In fact,
Kolmogorov--Smirnov test indicates the probability of no variability in
the light curve is 0.0012.

\section{Discussion}
\subsection{Thermal Component}
We have calculated the electron number density and the age of the
diffuse thermal plasma of region 1 and 2 on the basis of the best-fit
parameters summarized in table~\ref{tab:param_1} and
\ref{tab:param_2}. We assume that the plasma in region~1 distributes
uniformly within a sphere with a radius of $\timeform{1'.4}$, which is
the Half-Width-of-Half-Maximum (HWHM) radius obtained from the Chandra
image \citep{2008arXiv0803.0682H}. Assuming the distance to CTB37B of
10.2~kpc \citep{1975A&A....45..239C},
we obtain the real radius of region 1 to be $r_{\rm
reg1}=1.3\times10^{19}$~cm. Accordingly, the volume of the region~1
plasma is $V_{\rm reg1}=(4/3)\pi r_{\rm reg1}^3 =
9.2\times10^{57}$~cm$^{3}$. In the same way, from HWHM of an estimated
image size of $\timeform{1'.9}\times\timeform{0'.9}$, the real length of
semi-major and semi-minor axes result in $r_{\rm
reg2,l}=1.7\times10^{19}$~cm and $r_{\rm reg2,s}=8.2\times10^{18}$~cm,
respectively. Assuming that the line-of-sight extent of region 2 is
$r_{\rm reg2,s}$, the volume of region 2 is calculated to be $V_{\rm
reg2}=(4/3)\pi r_{\rm reg2,l} r_{\rm reg2,s}^2 =
4.8\times10^{57}$~cm$^{3}$ (the resultant electron density becomes
smaller by a factor of $\sqrt2$ if we substitute $r_{\rm reg2,l}$ for
$r_{\rm reg2,s}$). With the aid of the emission measures obtained from
the spectral fitting, $E.M.$ $= \int
n_{e}n_{H}dV=2.1^{+1.6}_{-1.0}\times10^{58}~\rm{cm^{-3}}$~(region 1) and
$2.6\times10^{57}~\rm{cm^{-3}}$~(region 2 upper limit,
see \S~4.2), the electron number density of region
1 and 2 are
\[
 n_{\rm e,\,reg1}\;=\; 1.7\;(1.2-2.2) \;\; \rm{[cm^{-3}]}  \nonumber
\]
and
\[
 n_{\rm e,\,reg2}\;\le\; 0.82 \;\; \rm{[cm^{-3}]}, \nonumber 
\]
where we adopt the relation $n_{\rm e} \approx
1.24~n_{\rm H}$ for fully ionized plasma. The parameter region in
parenthesis is those allowed at the 90\% confidence
level. Assuming the strong shock, we obtain the
pre-shock densities to be $\sim$0.43~cm$^3$ and
$\lesssim$0.21~cm$^3$, respectively, which are
significantly lower than the average density of the interstellar matter
in the galactic plane. That means CTB37B exploded in a
low density space. The density of region 1 together
with the ionization parameter obtained from the fit of the region 1
spectra $n_{\rm e,\,reg1}t=3.5^{+13}_{-1.1}\times10^{10}$ [cm$^{-3}$~s],
enables us to estimate the age of the plasma observationally for the
first time as
\[
 t_{\rm reg1}\;=\; 6.5\;(3.7-31) \times 10^2 \;\; \rm{[yr]}.
\]
CTB37B is one of the best candidates of SN393 in Chinese historical
record \citep{stephenson_2002}. The plasma age calculated from the
observed ionization parameter and emission measure supports this
identification.

The number of electrons ($N_{\rm e}=n_{\rm e}V$) in region 1 and 2 are
$N_{\rm e,\,reg1}=1.6\,(1.1-2.0)\times10^{58}$ and $N_{\rm
e,\,reg2}\le3.9\times10^{57}$. As a result, the total mass
included in the two regions are
15~(11--20)$M_\odot$ and $\le$3.7$M_\odot$, respectively, and the
thermal energy ($E = \frac{3}{2}(N_{\rm e}+N_{\rm H}+N_{\rm He})kT$) are
\begin{eqnarray}
E_{\rm reg1} &=& 6.4\;(4.1-8.6)\times10^{49} \;\;\rm{[ergs]}
 \nonumber \\
E_{\rm reg2} &\le& 1.6\times10^{49} \;\;\rm{[ergs]}.
 \nonumber
\end{eqnarray} 
under the assumption of energy equipartition between electrons and ions.
The total thermal energy could be larger if other
portions of the remnant are included and if the proton temperature is
significantly larger than the electron temperature as is expected for
supernova remnants with large shock velocities \citep{2007ApJ...654L..69G}.

\subsection{The Nature of the Point Sources}
\subsubsection{Source~A}
The best-fit spectral parameters of Chandra source~A and
those of Suzaku region~1 (source A
and the diffuse thermal emission) are summarized in
table~\ref{tab:param_1}. The measured hydrogen column densities
($N\rm{_{H}} \simeq 4\times 10^{22}$~cm$^{-2}$) are the same between
Chandra and Suzaku. However, since the Suzaku spectra below $\sim$1~keV
is dominated by the diffuse thermal component
(Fig.~\ref{fig:region1_spec_fit}), $N_{\rm H}$ obtained with Suzaku is
determined mainly by the diffuse emission.  This implies that source~A
is probably associated physically with the diffuse thermal emission in
region~1, and is most likely a neutron star (or a black hole) born with
the supernova explosion leading to CTB37B. In fact, the large photon
index of 3.2 and luminosity of 2.2$\times10^{34}\rm{ergs~s^{-1}}$ are
both consistent with those of anomalous X-ray pulsar (AXP;
\cite{1981Natur.293..202F,2006ApJ...645..556K}). Young
age of $\sim$700~yr~(\S~6.1) also supports this interpretation
({\it e.g.} the AXP 1E1841-045 is associated with
Kes~73 whose age is estimated to be 500-1000~yr;
\cite{2008ApJ...677..292T}).  Since the spin period of AXP is in the
range 6--12~s, it is natural that we have found no evidence of pulsation
from the Chandra data, because the frame time of the ACIS-I is 2.3~s. In
addition, the flux of the power law measured by Suzaku is apparently
greater that that with Chandra by a factor of 1.8 (\S4.1,
table~\ref{tab:param_1}). Note that the point spread function of the
Suzaku XRTs is not so sharp. Hence one may doubt that part of the
power-law flux detected by Suzaku can be attributed to a putative
diffuse non-thermal emission. However, $\sim$80\% of the Chandra
region~1 photons above 3~keV, where the power-law component is dominant
in the Suzaku spectra (see Fig.~\ref{fig:region1_spec_fit}), originates
from source~A (table~\ref{tab:source_count}). Since the diffuse emission
occupies only a small fraction in region~1, the flux difference between
Chandra and Suzaku cannot be explained unless source~A
has really varied. This kind of long term variation
is detected also from some other AXPs
\citep{2002ApJ...567.1067G,2003ApJ...588L..93K}. We therefore conclude
based on the Chandra and Suzaku observations that source~A is probably a
new AXP. We need a fast timing observation with an imaging detector with
a time resolution of less than 1~s to confirm our conclusion.

\subsubsection{Source~B}
From the spectrum fitting, The hydrogen column density obtained from the
fit to the Suzaku region~3 spectra $N_{\rm H} < 4\times 10^{20}{\rm
cm^{-2}}$ (table~\ref{tab:param_3}) is much smaller than that obtained
from region 1 and 2. This result indicates that source~B is a
foreground source. The best-fit plasma temperature of $kT \simeq
0.5$~keV is reminiscent of an active star. The existence of
flickering~(\S5.2) supports this suggestion.

\subsection{Non-Thermal Component}
X-ray emission from CTB37B is composed of the diffuse thermal component
(region 1) and the non-thermal component (region 2) as well as a point
source (source~A), as demonstrated in \S3 and \S4. Hence, CTB37B now
is the third SNR after RCW86 and Cas~A that possesses the thermal and non-thermal
X-ray emissions and TeV $\gamma$-ray emission all together. In addition,
CTB37B is now the fifth SNR, following RCW86, Cas~A, RX~J1713.7--3946 and Vela
Jr., from which non-thermal radiation is detected both in X-ray and TeV
$\gamma$-ray bands. The fluxes of the non-thermal emission of these five
non-thermal SNRs are compared in table~\ref{tab:nonthermal_snrs}.
 \begin{table}
 \begin{center}
 \caption{Comparison of non-thermal component with other SNRs.}
 \label{tab:nonthermal_snrs}
 \begin{tabular}{lcccccc}
   \hline\hline
   \multicolumn{1}{c}{Target Name}       & $L_x$ \footnotemark[$*$] &
   $L_{TeV}$ \footnotemark[$\dagger$] &
   $L_{TeV}$/$L_x$ & $\Gamma_x$ & $\Gamma_{TeV}$ & References \footnotemark[$\ddagger$]\\ 
   \hline
CTB37B~(region 2) & 0.97  & 0.59 \footnotemark[$\S$] &  0.61 & 1.5 & 2.3 & (1)\\
RCW86                 & 3.8    & 0.55 & 0.14 & 3.1 & 2.5 & (2)(3)\\
Cas A                 & 110 & 0.21 & 0.0019 & 3.1 & 2.4 & (4)(5) \\
RXJ1713               & 6.5   & 0.42     & 0.06     & 2.4 & 2.2 & (6)(7)\\
VELA Jr.              & $\sim$1.5$\times10^{-2}$ & 0.033  & $\sim$2.2 &
 2.6 & 2.1 & (8)\\
SN1006                & 2.1   & \textless0.15 & \textless0.1 & 2.7 & $\cdots$  & (9)(10)\\\hline\hline
   \multicolumn{7}{@{}l@{}}{\hbox to 0pt{\parbox{125mm}{\footnotesize
       \footnotemark[$*$] Unabsorbed flux in the
 2--10~keV band in the unit of 10$^{34}$~ergs~s$^{-1}$.\\
       \footnotemark[$\dagger$] Unabsorbed flux in the
 1--10~TeV band in the unit of 10$^{34}$~ergs~s$^{-1}$.\\
       \footnotemark[$\ddagger$] (1)~Aharonian et al. (2006); (2)~Bamba et
 al. (2000); (3)~Hoppe et al. (2007); (4)~Helder et al. (2008);
  (5)~Albert et al. (2007); (6)~Slane et al. (1999); (7)~Aharonian et al. (2004);
 (8)~Aharonian et al. (2005); (9)~Bamba et al. (2008); (10)~Aharonian et
   al. (2005b)\\ 
       \footnotemark[$\S$] TeV $\gamma$-ray flux contributed to region 2
  was calculated as a quarter of the total flux.
}\hss}}
   \end{tabular}
 \end{center}
 \end{table}
The non-thermal diffuse component detected from region 2 has a
remarkably flat X-ray spectrum with a photon index of
1.5. Since this photon index is equal to the typical radio photon index
(energy index $\alpha = 0.5$), the non-thermal emission
spectrum can be considered as extending from the radio band smoothly to
the X-ray band, thereby the roll-off energy results in as high as
$\gtsim$15~keV (table~\ref{tab:param_2}). 
This roll-off energy is higher than any other
SNR that is accompanied by the non-thermal X-ray and TeV $\gamma$-ray
emission, such as $\lesssim$9~keV for RX~J1713.7--3946
\citep{2008PASJ...60S.131T}, $\sim$0.23~keV for SN1006
\citep{2008PASJ...60S.153B}, $\sim$0.87~keV for RCW~86
\citep{2005ApJ...621..793B}.
This indicates high electron acceleration efficiency in
region~2. In addition,
the density around region 2 is considered as being
lower than in region 1, given that only the upper limit of the thermal
emission is obtained (\S~6.1). The lower density may indicate higher
shock velocity due to the smaller deceleration, which is consistent with
the fact that the roll-off frequency is proportional to the
square of the shock velocity~\citep{1999A&A...351..330A}.

It should be a matter of debate whether the TeV
$\gamma$-ray emission and the non-thermal X-ray emission from region~2
are produced by the same population of electrons, since the images of
these two bands shown in Fig.~\ref{fig:image_other} are far from similar
at first sight. The TeV $\gamma$-ray image is compatible with a shell
with a radius of $\sim$4$'$--6$'$ due to limited spatial resolution of
H.E.S.S. \citep{2008arXiv0803.0682H}, which is compatible with the size
of the radio shell. We thus assume that the TeV $\gamma$-ray emission
is powered through 1-Zone Inverse Compton scattering (IC) of the cosmic
microwave background due to the accelerated electrons.

\begin{figure}[htpb]
 \begin{center}
\FigureFile(80mm,50mm){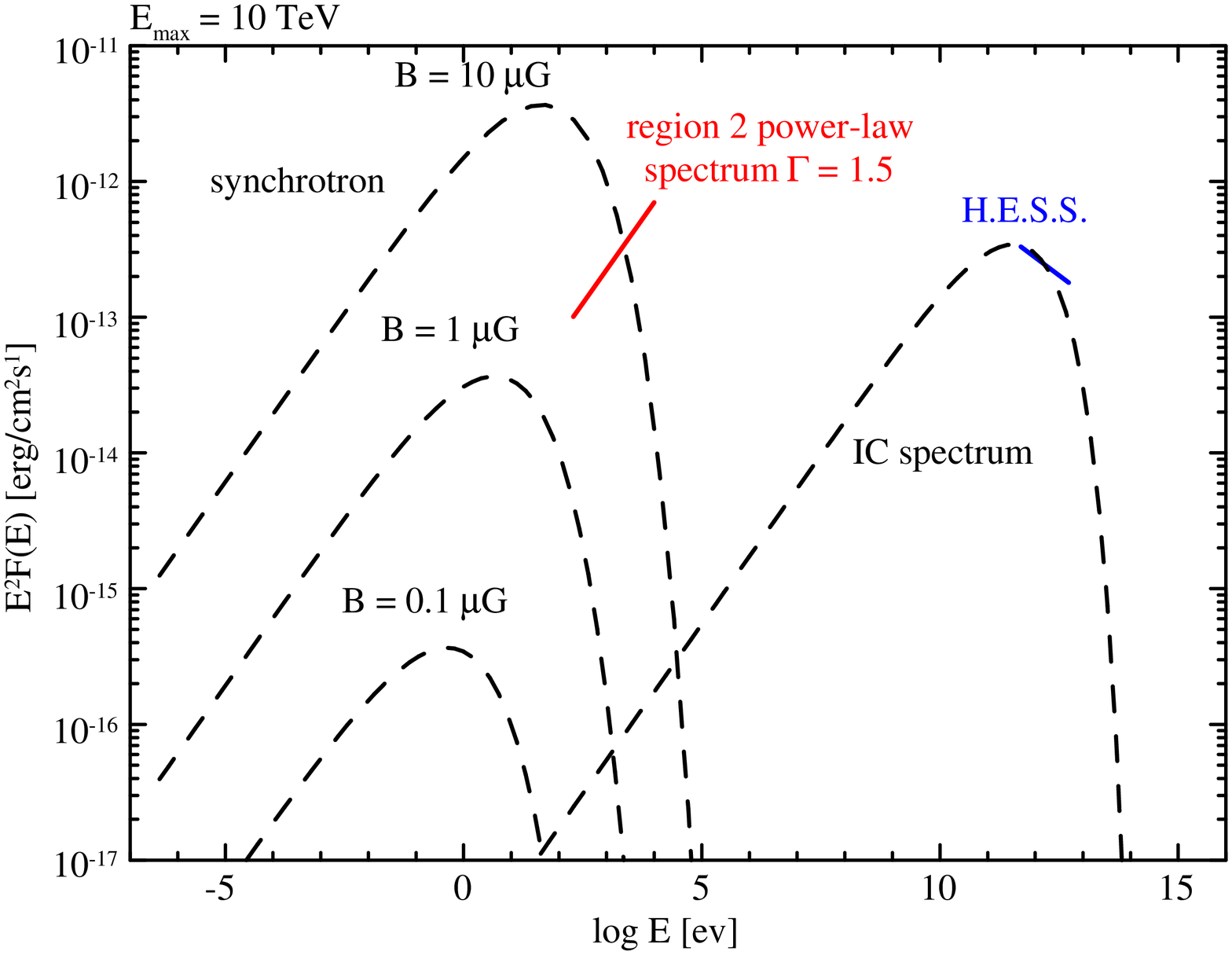}
 \end{center}
\caption{Spectrum energy distribution of region 2 from the X-ray and TeV
 $\gamma$-ray bands. The dashed plots on the left are
 the synchrotron radiation models with various magnetic field, and on
 the right is IC model spectrum. All the curves assume
 the maximum energy of electrons of 10~TeV. Blue line is the spectrum
 from H.E.S.S observation ($\Gamma = 2.3$; \cite{2006ApJ...636..777A}),
 and the red line is the Suzaku
 region 2 spectrum which is a power law with $\Gamma =
 1.5$.}  
\label{fig:E2F_spec}
\end{figure}

The maximum electron energy ($E_{\rm max}$) 
can be evaluated by the shape of TeV $\gamma$-ray
spectrum. Using the
H.E.S.S. spectrum whose photon index is 2.3
\citep{2006ApJ...636..777A} shown in
Fig.~\ref{fig:E2F_spec} in blue, we determined $E_{\rm max}$ of 10~TeV.
The red line in Fig.~\ref{fig:E2F_spec}, on the other
hand, is the X-ray power-law spectrum of region 2
with a photon index of 1.5. A series of the dashed
plots are the model spectra calculated under the
assumptions of $E_{\rm max}$ of 10~TeV, an index of the electron energy
distribution of 2, and various magnetic field (0.1, 1.0 and 10.0~${\rm
\mu}$G). It is clear from this figure that 1-zone IC
model is unable to explain the synchrotron X-ray
spectrum with any magnetic field strength. This
suggests that TeV $\gamma$-ray emission is
due to multi-zone IC scattering, or the decay of
neutral pions generated by the high energy proton
impacts.

\section{Conclusion}

We have obtained with Suzaku the images and the high quality spectra of
the supernova remnant CTB37B. The X-ray diffuse emission region
coincides with that of radio and TeV $\gamma$-ray. The X-ray emission
consists of thermal and non-thermal diffuse components
as well as a point source resolved by Chandra. CTB37B is the third SNR
from which thermal and non-thermal X-ray emissions as well as TeV
$\gamma$-ray emission are detected all together, and the fifth SNR that
is accompanied by non-thermal emission both in X-ray and TeV
$\gamma$-ray bands.

The diffuse thermal emission can be best described by a non-equilibrium
collisional ionization plasma model (NEI model) with a
temperature, an ionization parameter~($n_{\rm e}t$~[cm$^{-3}$s]), and
the abundances of 0.9$\pm0.2$~keV, 3.5~$^{+13}_{-1.1}\times10^{10}$, and
$\simeq$ 0.5$Z_{\odot}$~(Mg, Si), respectively. The
image size and the observed emission measure provides the number density
of the thermal electrons before the shock to be 0.2--0.4~cm$^{-3}$,
which is significantly lower than that of the Galactic plane. This
suggests that the supernova explosion associated with CTB37B took place
at a low density space.
From the ionization parameter and the number density of the thermal
electron, the age of the plasma is found to be
$\sim$650~$^{+2500}_{-300}$~yr. This is consistent with the tentative
identification of CTB37B with SN393 within the error.

In contrast, the diffuse component occupying southern part of CTB37B
(region 2) is non-thermal and represented by a
power-law model or a srcut model. The photon index of 1.5 is
significantly smaller than any other non-thermal SNR,
but is consistent with that of a typical non-thermal SNR in the radio
band. The srcut model fit with its normalization set free to vary
therefore results in a high roll-off energy of $>$15~keV.
Under the assumption that TeV
$\gamma$-ray was emitted by 1-zone IC scattering, there are no solution
for magnetic field strength that can reproduce the
observed synchrotron spectrum in
X-ray. This suggests that TeV $\gamma$-ray is
produced by multi-zone IC scattering, or by the decay
of neutral pions generated by the high energy
proton impacts.

Owing to the high spatial resolution of Chandra, a point source is
resolved from the brightest part of the Suzaku image of CTB37B
(region~1). Its association to the diffuse thermal emission indicated by
$N_{\rm H}$, the photon index of $\sim$3, the X-ray luminosity of order
$10^{34}$~erg~s$^{-1}$, and the long term flux variation evident from
the Chandra and Suzaku observations all indicate that the point source
is a new anomalous X-ray pulsar. A high speed
photometric observation is encouraged.

\bigskip
The authors are greatful to all of the Suzaku and H.E.S.S. team
members. We also thank H. Yamaguchi, T. Tanaka and J. Vink for useful comments.
This work was supported in part by Grant-in-Aid for Scientific Research
of the Japanese Ministry of Education, Culture, Sports, Science and
Technology, No.~19$\cdot$4014 (A.~B.), No.~18740153, No.~19047004 (R.~Y.).


\begin{thebibliography}{}

\bibitem[Aharonian et al.(1997)]{aharonian_1997} Aharonian, F.~A., 
Atoyan, A.~M., \& Kifune, T.\ 1997, \mnras, 291, 162 

\bibitem[Aharonian \& Atoyan(1999)]{1999A&A...351..330A} Aharonian, F.~A., \& Atoyan, A.~M. 1999, \aap, 351, 330 


\bibitem[Aharonian et al.(2004)]{aharonian_2004} Aharonian, F.~A., et 
al.\ 2004, \nat, 432, 75 

\bibitem[Aharonian et al.(2005)]{aharonian_2005a} Aharonian, F., et 
al.\ 2005, \aap, 437, L7 

\bibitem[Aharonian et al.(2005)]{aharonian_2005b} Aharonian, F., et 
al.\ 2005, \aap, 437, 135 

\bibitem[Aharonian et al.(2006)]{2006ApJ...636..777A} Aharonian, F., et 
al.\ 2006, \apj, 636, 777 

\bibitem[Aharonian et al.(2007)]{2007A&A...472..489A} Aharonian, F., et 
al.\ 2007, \aap, 472, 489 

\bibitem[Aharonian et al.(2008)]{2008arXiv0803.0682H} HESS 
Collaboration: F.~Aharonian 2008, ArXiv e-prints, 803, arXiv:0803.0682  

\bibitem[Albert et al.(2007)]{2007A&A...474..937A} Albert, J., et al.\
				2007, \aap, 474, 937

\bibitem[Anders \& Grevesse(1989)]{1989GeCoA..53..197A} Anders, E., \& 
Grevesse, N.\ 1989, \gca, 53, 197 

\bibitem[Bamba et al.(2000)]{2000PASJ...52.1157B} Bamba, A., Koyama, K., \& 
Tomida, H.\ 2000, \pasj, 52, 1157 

\bibitem[Bamba et al.(2005)]{2005ApJ...621..793B} Bamba, A., Yamazaki, R., 
Yoshida, T., Terasawa, T., \& Koyama, K.\ 2005, \apj, 621, 793 

\bibitem[Bamba et al.(2008)]{2008PASJ...60S.153B} Bamba, A., et al.\ 2008, 
\pasj, 60, 153 

\bibitem[Borkowski et al.(1994)]{1994ApJ...429..710B} Borkowski, K.~J., 
Sarazin, C.~L., \& Blondin, J.~M.\ 1994, \apj, 429, 710 

\bibitem[Borkowski et al.(2001)]{2001ApJ...548..820B} Borkowski, K.~J., 
Lyerly, W.~J., \& Reynolds, S.~P.\ 2001, \apj, 548, 820 

\bibitem[Case \& Bhattacharya(1998)]{1998ApJ...504..761C} Case, G.~L.,
\& Bhattacharya, D.\ 1998, \apj, 504, 761

\bibitem[Caswell et al.(1975)]{1975A&A....45..239C} Caswell, J.~L., Murray, 
J.~D., Roger, R.~S., Cole, D.~J., \& Cooke, D.~J.\ 1975, \aap, 45, 239 

\bibitem[Condon et al.(1998)]{1998AJ....115.1693C} Condon, J.~J., Cotton, 
W.~D., Greisen, E.~W., Yin, Q.~F., Perley, R.~A., Taylor, G.~B., 
\& Broderick, J.~J.\ 1998, \aj, 115, 1693 

\bibitem[Fahlman
\& Gregory(1981)]{1981Natur.293..202F} Fahlman, G.~G., \& Gregory,  
P.~C.\ 1981, \nat, 293, 202

\bibitem[Frail et al.(1996)]{1996AJ....111.1651F} Frail, D.~A., Goss, 
W.~M., Reynoso, E.~M., Giacani, E.~B., Green, A.~J., \& Otrupcek, R.\ 1996, 
\aj, 111, 1651

\bibitem[Gavriil \& Kaspi(2002)]{2002ApJ...567.1067G} Gavriil, F.~P., \& Kaspi, V.~M.\ 2002, \apj, 567, 1067 

\bibitem[Ghavamian et al.(2007)]{2007ApJ...654L..69G} Ghavamian, P., 
Laming, J.~M., \& Rakowski, C.~E.\ 2007, \apjl, 654, L69 

\bibitem[Hamilton et al.(1983)]{1983ApJS...51..115H} Hamilton, A.~J.~S., 
Chevalier, R.~A., \& Sarazin, C.~L.\ 1983, \apjs, 51, 115 

\bibitem[Helder 
\& Vink(2008)]{2008arXiv0806.3748H} Helder, E.~A., \& Vink, J.\
				2008, ArXiv e-prints, 806,
				arXiv:0806.3748

\bibitem[Hoppe et al.(2007)]{hoppe_2007} Hoppe, S., 
Lemoine-Goumard, M., \& for the H.~E.~S.~S.~Collaboration 2007, ArXiv 
e-prints, 709, arXiv:0709.4103 

\bibitem[Ishisaki et al.(2007)]{2007PASJ...59S.113I} Ishisaki, Y., et al.\ 
2007, \pasj, 59, 113 

\bibitem[Kaastra et al.(1996)]{1996uxsa.conf..411K} Kaastra, J.~S., Mewe, R., 
\& Nieuwenhuijzen, H.\ 1996, UV and X-ray Spectroscopy of Astrophysical and Laboratory Plasmas, 411 

\bibitem[Kaspi et al.(2003)]{2003ApJ...588L..93K} Kaspi, V.~M., Gavriil, 
F.~P., Woods, P.~M., Jensen, J.~B., Roberts, M.~S.~E., 
\& Chakrabarty, D.\ 2003, \apjl, 588, L93 

\bibitem[Kassim et al.(1991)]{1991ApJ...374..212K} Kassim, N.~E., Weiler, 
K.~W., \& Baum, S.~A.\ 1991, \apj, 374, 212 

\bibitem[Kokubun et al.(2007)]{2007PASJ...59S..53K} Kokubun, M., et al.\ 
2007, \pasj, 59, 53 

\bibitem[Koyama et al.(1995)]{1995Natur.378..255K} Koyama, K., Petre, R., 
Gotthelf, E.~V., Hwang, U., Matsuura, M., Ozaki, M., \& Holt, S.~S.\ 1995, 
\nat, 378, 255 

\bibitem[Koyama et al.(1997)]{1997PASJ...49L...7K} Koyama, K., Kinugasa, 
K., Matsuzaki, K., Nishiuchi, M., Sugizaki, M., Torii, K., Yamauchi, S., \& 
Aschenbach, B.\ 1997, \pasj, 49, L7 

\bibitem[Koyama et al.(2007)]{2007PASJ...59S..23K} Koyama, K., et al.\ 
2007, \pasj, 59, 23 

\bibitem[Kuiper et al.(2006)]{2006ApJ...645..556K} Kuiper, L., Hermsen, W., 
den Hartog, P.~R., \& Collmar, W.\ 2006, \apj, 645, 556 

\bibitem[Liedahl et al.(1995)]{1995ApJ...438L.115L} Liedahl, D.~A., 
Osterheld, A.~L., \& Goldstein, W.~H.\ 1995, \apjl, 438, L115 

\bibitem[Makishima et al.(1996)]{1996PASJ...48..171M} Makishima, K., et 
al.\ 1996, \pasj, 48, 171 

\bibitem[Matsumoto et al.(2007)]{2007PASJ...59S.199M} Matsumoto, H., et 
al.\ 2007, \pasj, 59, 199 

\bibitem[Mewe et 
al.(1985)]{1985A&AS...62..197M} Mewe, R., Gronenschild, E.~H.~B.~M., \& van den Oord, G.~H.~J.\ 1985, \aaps, 62, 197 

\bibitem[Mewe et 
al.(1986)]{1986A&AS...65..511M} Mewe, R., Lemen, J.~R., \& van den Oord, G.~H.~J.\ 1986, \aaps, 65, 511 

\bibitem[Milne et al.(1979)]{1979MNRAS.188..437M} Milne, D.~K., Goss, 
W.~M., Haynes, R.~F., Wellington, K.~J., Caswell, J.~L., \& Skellern, 
D.~J.\ 1979, \mnras, 188, 437 

\bibitem[Mitsuda et al.(2007)]{2007PASJ...59S...1M} Mitsuda, K., et al.\ 
2007, \pasj, 59, 1 

\bibitem[Ohashi et al.(1996)]{1996PASJ...48..157O} Ohashi, T., et al.\ 
1996, \pasj, 48, 157 

\bibitem[Reynolds(1998)]{1998ApJ...493..375R} Reynolds, S.~P.\ 1998, \apj, 
493, 375 

\bibitem[Reynolds \& Keohane(1999)]{1999ApJ...525..368R} Reynolds, S.~P., 
\& Keohane, J.~W.\ 1999, \apj, 525, 368 

\bibitem[Reynoso \& Mangum(2000)]{2000ApJ...545..874R} Reynoso, E.~M.,
\& Mangum, J.~G.\ 2000, \apj, 545, 874 

\bibitem[Serlemitsos et al.(2007)]{2007PASJ...59S...9S} Serlemitsos, P.~J., 
et al.\ 2007, \pasj, 59, S9 

\bibitem[Slane et al.(1999)]{slane_1999} Slane, P., Gaensler, 
B.~M., Dame, T.~M., Hughes, J.~P., Plucinsky, P.~P., \& Green, A.\ 1999, 
\apj, 525, 357 

\bibitem[Smith 
\& Wang(2004)]{2004ApJ...611..881S} Smith, D.~A., \& Wang, Q.~D.\ 2004,
				\apj, 611, 881 

\bibitem[Stephenson \& Green(2002)]{stephenson_2002} Stephenson, F.~R., 
\& Green, D.~A.\ 2002, Unknown,  
 
\bibitem[Sugizaki et al.(2001)]{2001ApJS..134...77S} Sugizaki, M., Mitsuda, 
K., Kaneda, H., Matsuzaki, K., Yamauchi, S., \& Koyama, K.\ 2001, \apjs, 
134, 77 

\bibitem[Takahashi et al.(2007)]{2007PASJ...59S..35T} Takahashi, T., et 
al.\ 2007, \pasj, 59, 35 

\bibitem[Takahashi et al.(2008)]{2008PASJ...60S.131T} Takahashi, T., et 
al.\ 2008, \pasj, 60, 131 

\bibitem[Tanaka et al.(1994)]{1994PASJ...46L..37T} Tanaka, Y., Inoue, H., 
\& Holt, S.~S.\ 1994, \pasj, 46, L37 

\bibitem[Tian 
\& Leahy(2008)]{2008ApJ...677..292T} Tian, W.~W., \& Leahy, D.~A.\ 2008, \apj, 677, 292 

\bibitem[Voges et al.(1999)]{1999A&A...349..389V} Voges, W., et al.\ 1999, 
\aap, 349, 389 

\hspace{-5mm}Yamauchi, S., Ueno, M., Koyama, K., \& Bamba, A.\ 2008, \pasj, submitted

\end{thebibliography}
\end{document}